\documentclass[aps,prb,superscriptaddress,twocolumn,floatfix,citeautoscript]{revtex4-1}
\usepackage[pdftex]{graphicx}
\usepackage{color}
\usepackage{subfigure}
\usepackage{booktabs}
\usepackage{array}
\graphicspath{{./figures/}}

\newcommand{\ORNL}{Materials Science and Technology Division,
 Oak Ridge National Laboratory, Oak Ridge, Tennessee 37831-6138, USA}
 
\newcommand{\unigrenoble}{Univ. Grenoble Alpes, INAC-SP2M, L$\_{Sim}$, F-38000 Grenoble, France}
\newcommand{\ceagrenoble}{CEA, INAC-SP2M, Atomistic Simulation Lab., F-38000 Grenoble, France}

\begin{document}

\title{Strain effect and intermixing at the Si surface: A hybrid quantum and molecular mechanics study}


\author{Laurent Karim B\'{e}land}
\email[]{belandlk@ornl.gov}
\affiliation{Regroupement Qu\'{e}b\'{e}cois sur les Mat\'{e}riaux de Pointe(RQMP), \\
D\'{e}partement de physique, Universit\'{e} de Montr\'{e}al, \\
Case Postale 6128, Succursale Centre-ville, Montr\'{e}al, Qu\'{e}bec, H3C 3J7,Canada}
\affiliation{\ORNL}
\author{Eduardo Machado-Charry}
\affiliation{\unigrenoble}
\author{Pascal Pochet}
\email[]{pascal.pochet@cea.fr}
\affiliation{\unigrenoble}
\affiliation{\ceagrenoble}

\author{Normand Mousseau}
\email[]{normand.mousseau@umontreal.ca}

\affiliation{Laboratoire de Physique Th\'eorique de la Mati\`ere Condens\'ee, Universit\'e Pierre et Marie Curie, 4 Place Jussieu, F-75252 Paris Cedex 05, France}
\affiliation{Regroupement Qu\'{e}b\'{e}cois sur les Mat\'{e}riaux de Pointe(RQMP), \\
D\'{e}partement de physique, Universit\'{e} de Montr\'{e}al, \\
Case Postale 6128, Succursale Centre-ville, Montr\'{e}al, Qu\'{e}bec, H3C 3J7,Canada}

\date{\today}

\begin{abstract}

We investigate Ge mixing at the Si(001) surface and characterize the $2\times N$ Si(001) reconstruction by means of hybrid quantum and molecular mechanics calculations (QM/MM). Avoiding fake elastic dampening, this scheme allows to correctly take into account long range deformation induced by 
reconstruted and defective surfaces. We focus in particular on the dimer vacancy line (DVL) and its interaction with Ge adatoms. We first show that calculated formation energies for these defects are highly dependent on the choice of chemical potential and that the latter must be chosen carefully. Characterizing the effect of the DVL on the deformation field, we also find that the DVL favors Ge segregation in the fourth layer close to the DVL. Using the activation-relaxation technique (ART nouveau) and QM/MM, we show that a complex diffusion path permits the substitution of the Ge atom in the fourth layer, with barriers compatible with mixing observed at intermediate temperature.


\end{abstract}


\maketitle

The deposition of Ge on the Si(001) surface is a model system for Stransky-Krastanow
growth, a process of great technological relevance in microelectronics.
~\cite{2002RPPh...65...27B,Kiravittaya_adv}  This
process is known to be driven by several factors, such as lattice mismatch strain, surface
reconstruction, adatom diffusion and dimerization and inter-diffusion of Ge and Si. While
much attention has been focused on the island formation, a number of results point to the existence of complex interaction between the deposited and the substrate species in the wetting phase, both on the surface and deep
below.~\cite{liu1997effect,zhou2011atomic} 

With a 4.2~\% lattice mismatch, Ge atoms deposited on the Si(001) surface first
form a wetting layer that adopts the same $2\times 1$ reconstruction as the top Si layer.
The resulting compressive strain is partially accommodated in the wetting layer by
removing rows of dimers at regular intervals, forming a $2\times N$ periodic arrangement
of dimer vacancy lines (DVLs)~\cite{liu1997effect}. The exact interval is controlled by
the wetting layer thickness as well as the amount of intermixing, which also affects
internal strain. A number of numerical descriptions of the $2\times N$ reconstruction have been
reported, using empirical potentials
~\cite{nurminen2003comparative,nurminen2003reconstruction,ciobanu2004comparative},
tight-binding~\cite{li2003tight} and DFT
description~\cite{wang1993dimer,oviedo2002first,beck2004surface,varga2004critical}. These
studies find that DVLs are the favored arrangement for surface dimer vacancies, and
predict small (a few hundred meV per vacant dimer) to negative DVL formation energies,
even on unstrained Si(001). Recent work has shown that \emph{c}($2\times8$)
reconstruction can appear on the Si(111) surfaces to which a 0.03 tensile
strain is applied \cite{zhachuk2013strain}. Comparison of DVLs on strained Ge(001) and strained Si(001) was made
using a classical potential~\cite{ciobanu2004comparative}, but not using \emph{ab initio}
methods. To our knowledge, no \emph{ab initio} study differentiates pure strain/stress
effects from Ge/Si alloying and interface effects. 

A recent surface x-ray diffraction (SXRD) experiment~\cite{zhou2011atomic} has determined
the average atomic positions in the presence of a dimer vacancy line (DVL) and rekindled
interest for the effects of the DVL on elastic deformation and Ge intermixing. While some
comparisons have been made between the SXRD results ~\cite{zhou2011atomic} and a Monte
Carlo study \cite{nurminen2003reconstruction} based on the Stillinger-Weber potential
\cite{stillinger1985computer}, a more comprehensive atomistic description of the DVL
structure as well as its impact on Ge diffusion is still lacking.

Given its importance, the mixing of Si and Ge between adatoms, ad-dimers and the surface
dimers has been extensively investigated by both experiments
\cite{patthey1995mixed,nakajima1999direct,uberuaga2000diffusion,qin2000diffusional,nakajima2000intermixing,bussmann2010ge,zhou2011atomic} and theoretical calculations
\cite{uberuaga2000diffusion,ko1999ab,lu2000unique,lu2002mixed,nurminen2003reconstruction,wagner2004simulation,zipoli2008first}. It is now established experimentally and
theoretically that Ge can mix with the surface dimers at room temperature, and that deep
intermixing to the third and fourth atomic layers occurs at temperatures of 773~K and
higher. Calculations also showed kinetic paths for Si/Ge exchange at the (105) surface
\cite{cereda2010si}. Furthermore, SXRD \cite{zhou2011atomic} and classical Monte Carlo
\cite{nurminen2003reconstruction} studies find that more intermixing takes place far from
the DVL, interpreted as the consequence of compressive strain near the DVL.

Experiments~\cite{nakajima1999direct,sasaki1994auger,patthey1995mixed} suggest
that deep intermixing could happen at even lower temperatures, as low as 573~K, but the results are within the measurement's margins of error and further work is needed to clarify this issue. Interestingly, an empirical rule-of-thumb
in metallurgy~\cite{cahn1996physical} states that diffusion becomes important when temperatures reach one third of the
melting temperature. In the case of Si, this corresponds to 562K, which would indicate that
intermixing might be possible at this temperature. While thermodynamical computations \cite{uberuaga2000diffusion,cho2000ge} open the door for Ge fractional occupation of
a few percents in the fourth layer at a temperature of 600~K, no kinetic path permitting
significant intermixing at these temperatures has been found, leaving open the question
as to whether Ge could diffuse deep below the surface on an experimental time scale.

In this article, we use a quantum mechanical/molecular mechanics approach based on the
highly parallelizable DFT waveled-based BigDFT
package~\cite{genovese2008daubechies,genovese2011daubechies} to investigate the specific
role of strain on both the DVL and Ge diffusion into the bulk. More precisely, we first
look at the elastic, energetic and thermodynamical effects of the creation of a DVL in a
strained Si box. Coupling this package with the Activation-Relaxation Technique (ART
nouveau or ARTn)~\cite{malek2000dynamics}, we also explore various kinetic pathways that could
allow Ge to diffuse deep below the surface during deposition below 600~K. By focusing on
the strain effect, we provide clear numerical evidence as to the creation of DVL, the
Ge subsuface diffusion and its relation to the DVL.

\section{Methodology}

Most \emph{ab initio} studies of elastic deformations near surfaces are limited by the
depth of the slab used to simulate the system. Atoms at the bottom of the slab are
typically frozen, possibly resulting in important elastic dampening when the sample depth
is not sufficiently large. It is therefore essential to use simulation cells of
sufficient size to allow unconstrained elastic relaxation to take place near the surface.
This remains a challenge for fully quantum mechanical approaches as this means spending a
considerable amount of computer efforts on relatively trivial displacement away from the
zone of interest.

It is possible to reduce the computational costs in certain systems by using a hydrid
quantum mechanics/molecular mechanics approach (QM/MM) where atoms near the surface are
treated quantum-mechanically and those deep in the bulk with a much cheaper empirical
potential~\cite{warshel1976theoretical}. While such an approach can be used with any
quantum mechanical code (e.g. Ref. \onlinecite{PhysRevLett.105.185502}), it is particularly well suited for a local-basis implementation,
such as BigDFT, a powerful wavelet-based DFT package that we use here
~\cite{genovese2008daubechies,genovese2011daubechies}

In our implementation of the QM/MM approach, we use BigDFT with GGA/PBE exchange-correlations functionals to accurately describe the surface states. For the MM region, we select the original Stillinger-Weber potential~\cite{stillinger1985computer}, simply adapting the lattice parameter to the QM value of 5.465~\AA. This potential describes adequately the small harmonic displacements around the global minimum. Minimizations throughout this work are done using the FIRE algorithm \cite{bitzek2006structural}.

Throughout this study, we consider three Si(001) configurations: model 1, model 2 and model 3. All models are slab configurations where the first 8 atomic layers are described by BigDFT. The MM layers are placed below the QM layers and are described using the Stillinger-Weber potential. We detail these model configurations in table \ref{tab:models}.

\begin{table*}
\centering
\caption{Geometrical details of 
the three Si(001) model configurations used in this study. All models are slab configurations where the first 8 atomic layers are described by BigDFT. The molecular mechanical (MM) layers are placed below the quantum mechanical (QM) layers and are described using the Stillinger-Weber potential. The number of Si atoms in each of the two regions are indicated in the last two columns.
}
\begin{tabular}{|c|c|c|c|c|c|c|}
\hline
Model & Atoms per layer & Surface size & QM layers & MM layers & QM atoms & MM atoms\\ \hline
1 & 16 & 1.55 nm x 1.55 nm  & 8 & 20 & 128 & 320\\
2 & 24 & 1.55 nm x 2.32 nm  & 8 & 20 & 192 & 480\\
3 & 40 & 1.55 nm x 3.86 nm  & 8 & 36 & 320 & 1440\\
\hline
\end{tabular}

\label{tab:models}
\end{table*}

The interface between the QM region and the MM region is described by one layer of buffer atoms. These atoms are included in the QM calculation, but are moved only according to the MM forces on them. When a BigDFT calculation is launched, the bottom of the QM region is passivated with H atoms positioned at half the bond length between neighboring QM to ensure that no dangling bond at the bottom of the QM slab affects the computation. This distance is chosen so as to minimize H-H interactions when the lattice is deformed. Since it is not possible to separate the potential energy contribution of each atom using DFT calculations, we must include the atoms in the passivated surface in potential energy calculations.

The minimum thickness for the buffer region was established first by relaxing a 216-atom box with periodic conditions and then H-passivating the top and the bottom. Then we computed the forces on this system, without further relaxation. In such a context, Si atoms with a non-zero force correspond to the region which is affected by finite-size effects. We found that this effect is limited to the first layer of atoms. We counter-checked this result by adding a single Si interstitial atom to the system and comparing the forces on the surrounding atoms in the periodic-box case and in the H-passivated slab case. Once again, only the first layer of Si atoms was affected by our scheme.

The QM thickness of our $2\times1$-reconstructed Si(001) system was determined by computing the
force on a single Ge adatom as a function of the QM region depth. This procedure was
executed using purely QM systems, with the same number of atoms per layer as model 2,
testing diverse slab thicknesses. We found that the QM-energy converged with 8 atomic
layers.

One should note that the QM/MM framework is appropriate for IV-group semiconductors but that it would necessitate important modifications for use in systems with important charge transfer, such as ionic materials. In such systems, H-passivation would probably not be adequate to minimize electronic effects in the bottom layers of the QM region. However, adaptation to other systems of IV-group semiconductors with different geometries is possible.

\begin{figure}
\centering
\subfigure[Top view]{ 
 \includegraphics[width=0.48\textwidth]{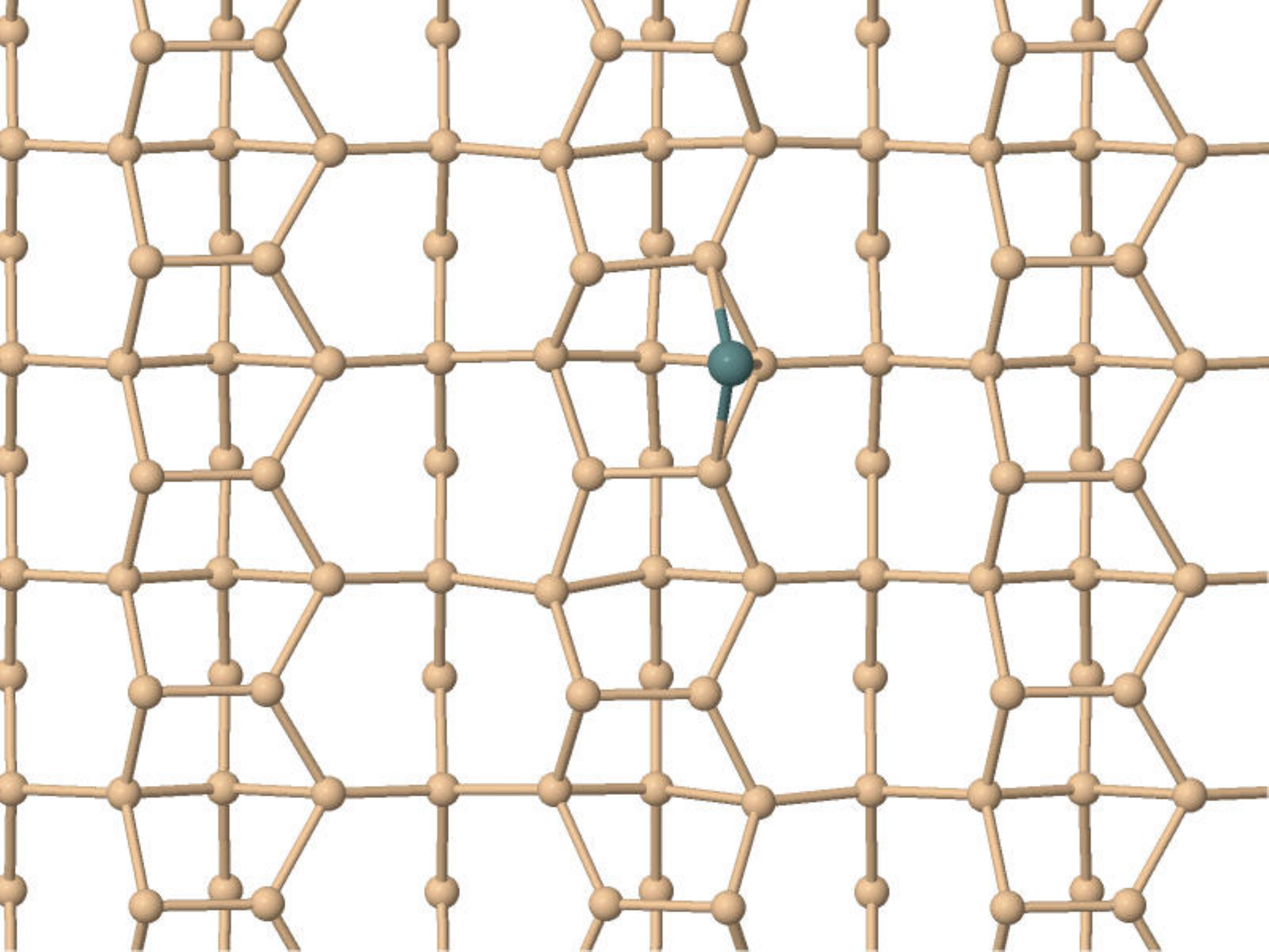}
 \label{fig:top_view}}
\subfigure[Side view]{ 
 \includegraphics[width=0.48\textwidth]{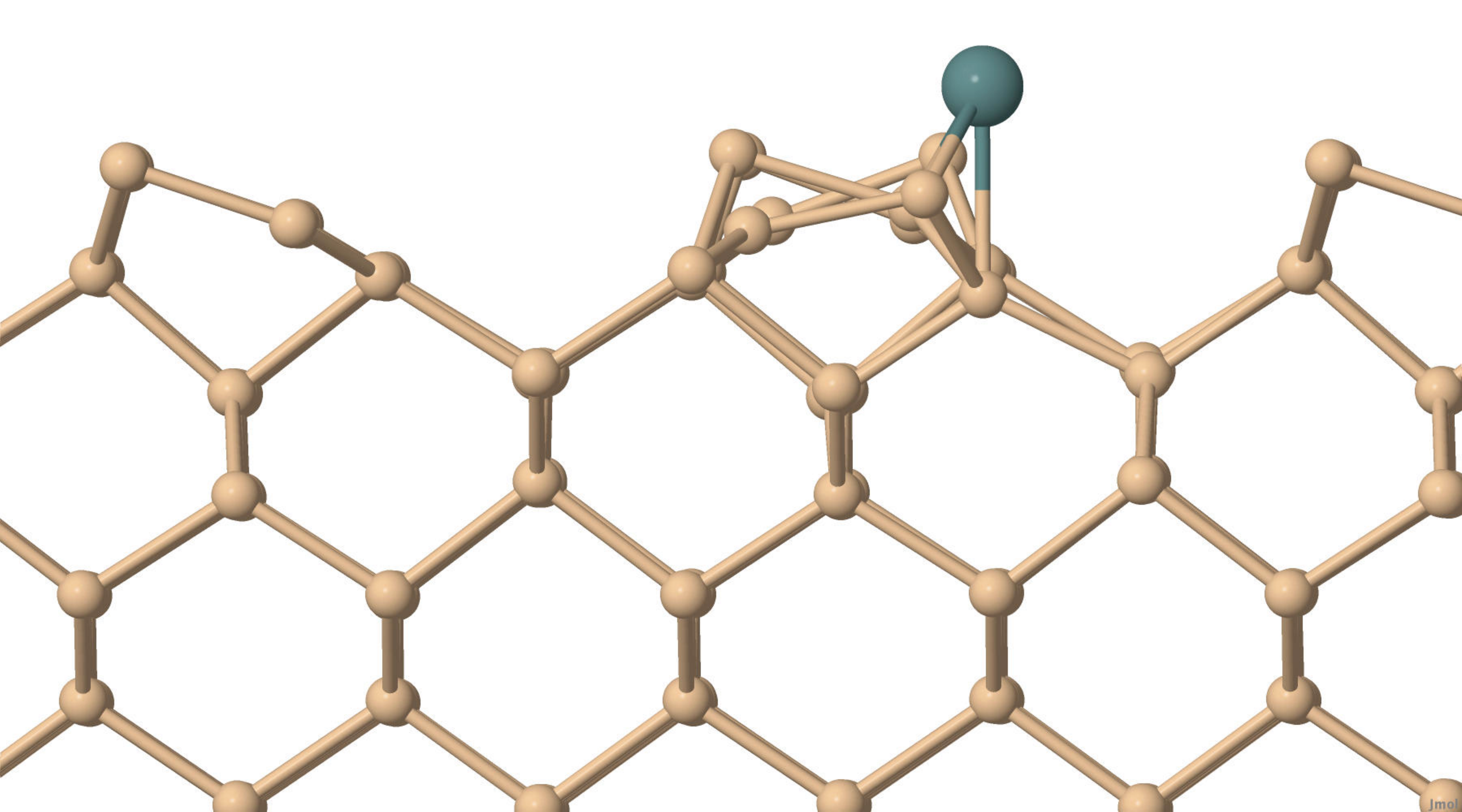}
 \label{fig:side_view}}
\caption{An illustration of a relaxed Si surface with a Ge adatom (in blue) in pedestal position. The surface is reconstructed in the so-called alternating fashion: one dimer row (on the left) is $2\times1$ reconstructed and the next (with the Ge adatom on top) is $2\times2$ reconstructed.} 
\label{fig:pedestal_alterning} 
\end{figure}

\subsection{The importance of long-range elastic effects}

To assess the need for a QM/MM scheme, we compute the energy and structural minima for
various Si(001) systems with different initial surface reconstructions. Computations on
variations of Model 1 are done with a $2\times1\times2$ Monkhorst-Pack grid. In the pure
QM case the bottom layer and H atoms are fixed. Our QM results agree with previous
studies: a 40 meV energy decrease per surface atom when switching from $2\times1$ to
$2\times2$, and a 2.5 meV energy decrease per surface atom when switching from $2\times2$
to $4\times2$. Using the QM/MM scheme leaves the same energy difference between
reconstructions, but the three relaxed structures ($2\times1$, $2\times2$, and
$4\times2$) show a DFT energy 0.625 meV per surface atom lower than in the pure QM case.
Although the bulk elastic properties affect the total energy, they have little
influence on the geometry of the reconstruction.

Starting from Model 2, we perform similar computations but with a Ge adatom in pedestal
position. The pedestal position, after minimization, is illustrated in figure
\ref{fig:pedestal_alterning}. We relax starting from three initial surface
reconstructions: $2\times1$, $2\times2$ and alternating $2\times1$ and $2\times2$ rows.In
the latter case, the Ge adatom was placed on a $2\times2$ row (see Fig.~
\ref{fig:pedestal_alterning}). We observe that the $2\times1$ reconstruction
spontaneously transforms to an alternating configuration after minimization in the
presence of a Ge atom.

In all these systems with a Ge adatom, the presence of a large MM region directly affects the energy minima at the top surface. In the $2\times1$ configuration, the final potential energy
per Si surface atom was 4 meV lower in the QM/MM case than the pure QM case. These
values, for the $2\times2$ and alternating dimers configurations, per Si surface atom, are 11 meV
and 4 meV, respectively. Clearly, therefore, correctly incorporation long-range elastic
deformations is necessary to obtain accurate total and relative surface energies.

\subsection{Modeling the Ge/Si surface reconstruction}

Dimer vacancy lines (DVL) appear after the deposition of a monolayer of Ge on top of
Si(001): in order to release compressive stress, some surface dimers will become vacant
and align themselves as vacancy lines. An illustration of such a configuration is given
in Fig.~\ref{fig:map_x}. Because of mixing, Ge concentrations in layers near the surface
may vary from zero to 100 percent, depending on deposition conditions. Full sampling of
these configurations is an expensive computational task (see, e.g.
Ref.~\onlinecite{nurminen2003reconstruction}) that is beyond our computational resources when doing
\emph{ab initio} calculations. We therefore consider two limiting cases: an unstrained
Si(001) substrate and a Si(001) substrate with 4\% biaxial compressive strain, mimicking
the lattice mismatch of Ge and Si. This approximation allows us to focus on strain and
stress effects of Ge/Si, leaving aside chemical alloying and interface effects.

\subsection{Energy landscape exploration}

For the exploration of the energy landscape of a single Ge atom added to the Si(001)
surface, we use a modified version of the bigDFT implementation of ARTn
\cite{PhysRevLett.77.4358,malek2000dynamics,machado2011optimized,mousseau2012activation}.
To accelerate convergence, we both follow the variable step procedure proposed by
Canc\'{e}s \emph{et al.} for the activation phase~\cite{cances2009some} and converge the
perpendicular direction with FIRE~\cite{bitzek2006structural}. ARTn is an efficient
open-ended unbiased search algorithm both for transition states. It has been used
successfully to characterize mixing in SiO$_2$~\cite{ganster2012first},
glasses~\cite{rodney2009distribution}, proteins~\cite{wei2002exploring}, defects in
iron~\cite{marinica2011energy} as well as study, \emph{ab initio}, diffusion in various
semiconductors \cite{el2006charge,levasseur2008ab,levasseur2008numerical}. ARTn
exploration were performed on Model 2 (24 atom/layer).

Some of the metastable states of great interest found by ARTn were discovered when
executing high barrier events. In some cases, we found new transition states by relaxing
pure-QM Nudged Elastic Bands \cite{jonsson1998nudged} (NEB) linking these states, as implemented in the BigDFT package.
These saddle points were then refined with the pure-QM ARTn followed by the QM/MM
ARTn. By this procedure, transition states with lower barriers were found.

\section{Characterization of the surface with a dimer vacancy line}

\begin{figure}
	\centering
    \includegraphics[width=9cm]{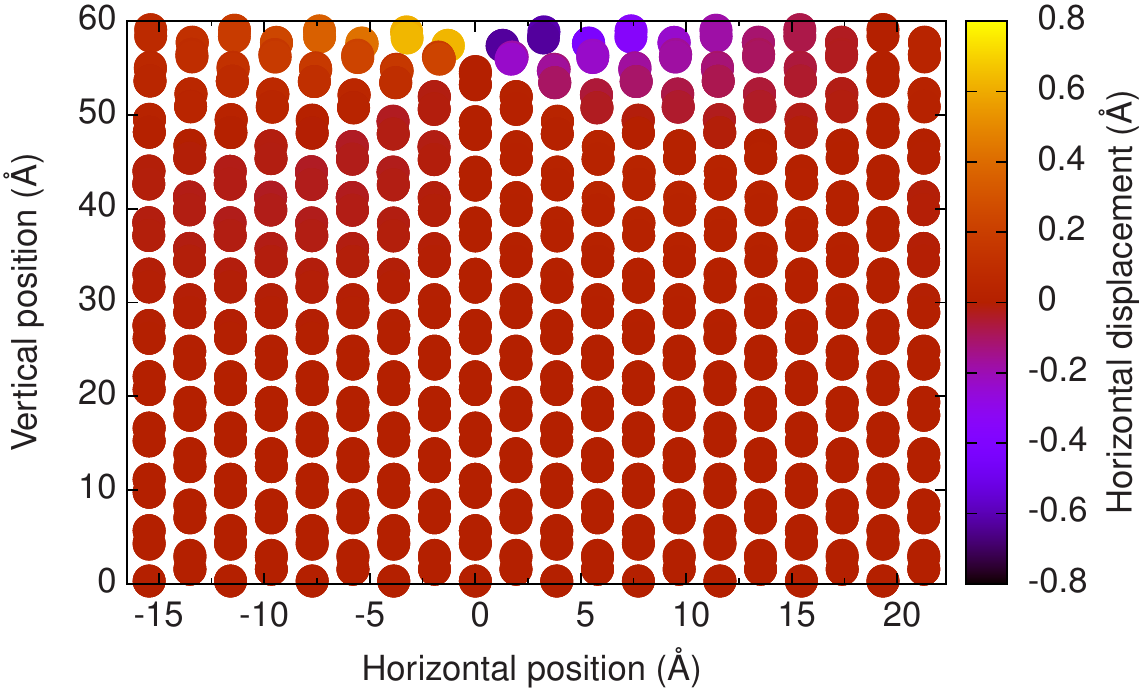}
    \caption{Horizontal displacements respective to the perfect $2\times1$ lattice in the presence of a DVL (missing dimer row near $x =0$~\AA ). Atoms with positive displacements are shifted towards the right and those with negative displacements towards the left.} 
\label{fig:map_x} 
\end{figure}

\begin{figure}
	\centering
    \includegraphics[width=9cm]{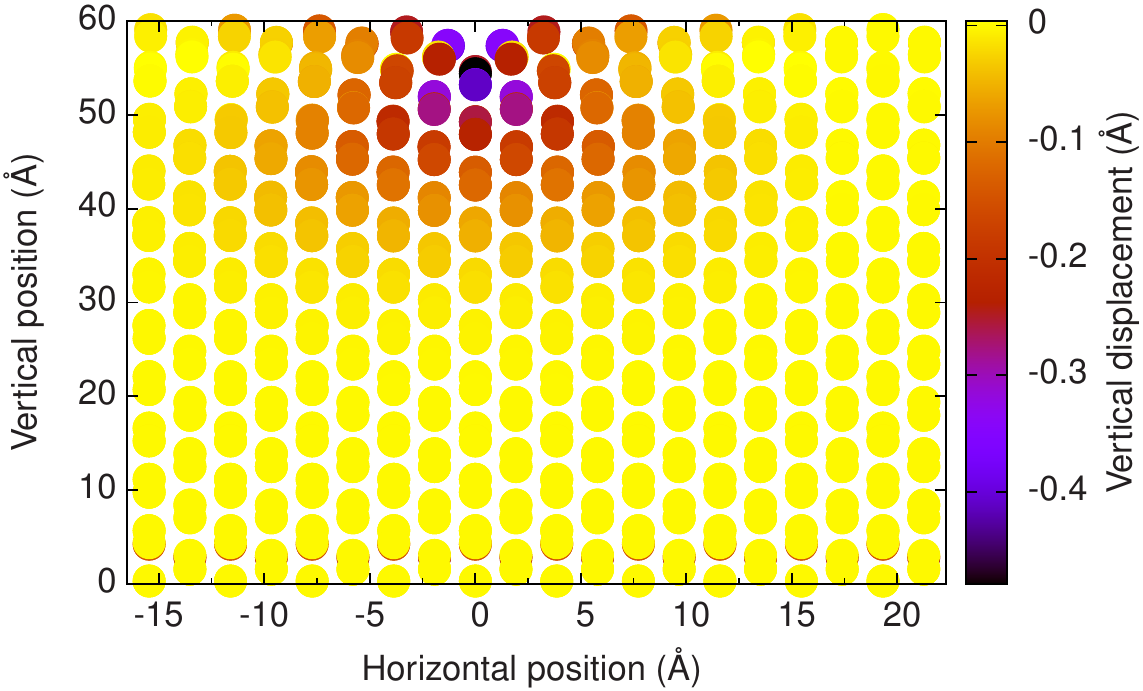}
    \caption{Vertical displacements respective to the perfect $2\times1$ lattice in the presence of a DVL (missing dimer row near $x =0$~\AA ). Atoms with negative displacements are shifted towards the bottom.} 
\label{fig:map_y} 
\end{figure}

The dimer vacancy line (DVL) is a surface defect characterized by one missing dimer out of every $N$ (with $N$ typically between 7 and 12, depending on the deposition conditions). Following conventional notation, we define the DVL as $2\times N$, the ``2'' referring to the dimer surface reconstruction. We select the $2\times10$ reconstruction, since $N=10$ sits approximatively half-way between the values found in the literature for high strain and low strain~\cite{nurminen2003reconstruction}.

As discussed above, the $2\times10$ reconstruction of Si was characterized using both the QM and the QM/MM methods. It is thought that the DVL is a consequence of the strain imposed on the Ge layer, since pure Ge(001) and Si(001) do not this defect.  This problem was studied using a classical potential \cite{ciobanu2004comparative} and we revisit it using an \emph{ab initio} method in Model 3, introducing two aligned dimer vacancies (the surface thus contains 18 dimers, spread on two rows). 

We first remove a dimer line for a configuration with a perfect $2\times1$ reconstruction and no strain. After relaxation, the system converges to a staggered surface configuration where atoms bordering the missing dimers sit at variable distances from the initial DVL, depending on whether they neighbored the top or bottom atom of a tilted surface dimer (the atoms neighboring the bottom dimer atom are 1.175 \AA~ from the DVL and the others are 2.425 \AA~ from the DVL). After repositioning these atoms at 1.175 \AA~ of the DVL, a new relaxation leads to a more stable state (300 meV per dimer vacancy compared to the metastable state), both with the QM and QM/MM techniques.

Although we obtain the same potential energy difference between the stable and metastable DVL configuration using the QM and QM/MM schemes, QM/MM leads to a potential energy lower than that of QM (300 meV per dimer vacancy), showing that the DVL causes long-range deformations in the bulk. The final configuration is illustrated in Fig.~ \ref{fig:map_x}. We used this configuration as a starting point for the minimization at 4~\% biaxial compressive strain, which was done using the QM/MM scheme.

\subsection{The formation energy of the DVL}

The DVL formation is associated with the removal of atoms. Its formation energy must therefore be computed in the grand canonical ensemble, which requires the chemical potential associated with the surface atoms. It can then be written as
\begin{equation}
  \label{eq:formation_energy}
E_f=E_{DVL} - E_{2\times1} + n_{vac}\mu_{dimer},
\end{equation}
where $E_{DVL}$ is the system's total energy with the DVL formed by removing $n_{vac}$ dimers, $E_{2\times1}$ is the total energy of the perfect $2\times1$ surface reconstructed simulation box and $\mu_{dimer}$ is the chemical potential associated with a reconstructed dimer. 

Previous studies used the bulk chemical potential as reference for the missing surface atoms forming the DVL. This choice is generally justified by considering that it is equivalent to using a sink and source of atoms at the edge of a terrace, since the displaced atom will cover what was previously a surface atom that becomes a bulk atom~\cite{ciobanu2004comparative,oviedo2002first,wang1993dimer}.

It is also possible to define a surface chemical potential as
\begin{equation}
  \label{eq:chem_dimer1}
\mu_{dimer}=\mu_{bulk} +  \frac{\gamma_{2\times1}}{n_{dimer}},
\end{equation}
where $\gamma_{2\times1}$ is the surface energy of a perfect $2\times 1$ reconstructed surface and  $n_{dimer}$ is the number of dimers on the $2\times 1$ reconstructed surface. This is equivalent to considering that the boundary is an infinite reservoir of surface dimers. Since our quantum system is H-passivated, we get:
\begin{equation}
  \label{eq:chem_dimer}
\mu_{dimer}=(E_{2\times1} -\gamma_{passivated}-n_{bulk}\mu_{bulk})/n_{dimer},
\end{equation}
where the H-passivated surface energy $\gamma_{passivated}$ is given by:
\begin{equation}
  \label{eq:surf_H}
\gamma_{passivated}=(E_{H-terminated} -n_{bulk}\mu_{bulk})/2.
\end{equation}
and $\mu_{bulk}$ is computed using a 216 atom box with periodic boundary conditions. $\gamma_{passivated}$ is computed using a system with 8 Si layers and two H-terminated (silicon passivated with hydrogen) surfaces.

These quantities are computed for 0~\% and 4~\% biaxial strain. Since we are using a slab
configuration, $\mu_{bulk}$ must account for the fact that the system can relax
vertically when under biaxial strain. Thus, the vertical size of the periodic box used to
compute the bulk chemical potential under compressive strain is adjusted using the
experimental 0.22 Poisson ratio of Si.

We report the formation energies $E_f$ for this slab with and without compressive strain
in Table~\ref{tab:DVL_formation_energy}. Since the surface cohesive energy is less than
that of the bulk, the DVL formation energies are shifted to higher values when using
$\mu_{dimer}$ rather than $\mu_{bulk}$, showing the importance of selecting the correct reference state. 

Indeed, while the DVL is formed at no cost in the unstrained silicon sample when using
the bulk chemical potential, the formation energy computed with respect to a surface
chemical potential suggests rather that the DVL is
unstable at zero pressure. For a box under compressive strain, DFT results using either
chemical potentials suggest a stable DVL. Interestingly, the Stillinger-Weber potential
shows that the DVL is stable if one uses the bulk chemical potential, but unstable if one
uses the reconstructed surface chemical potential. This result is not unexpected, since
this classical potential is well-known to predict bulk properties better than surface
properties.

Experiments systematically show the presence of vacant dimers but not DVLs on the
unstrained Si(001) surface. They are formed when cleaning the surface and survive
annealing \cite{zandvliet1997ordering}. In one study \cite{koo1995dimer}, when carefully
avoiding metal contamination, 1.7 \% of dimers are vacant, forming mostly single
vacancies, a percentage independent of the annealing temperature. This indicates that
vacant dimers are caused by a mechanical effect (surface cleaning) and not a
thermodynamical effect. In another study \cite{PhysRevLett.64.2406}, 9 \% of dimers are
vacant, forming small clusters.  No DVLs are observed. 

This contradicts previous computations that suggest that DVLs are more stable than
individual dimer vacancies, i.e. dimer vacancies tend to align
\cite{nurminen2003comparative,ciobanu2004comparative,li2003tight,oviedo2002first,beck2004s
 urface,varga2004critical}. 
%

As shown in Table~\ref{tab:DVL_formation_energy}, we resolve this paradox by demonstrating the importance of selecting the right chemical potential. Clearly the surface chemical potential and not the bulk chemical potential, generally favored, is the right choice. 

\begin{table}
\centering
\caption{
DFT and Stillinger-Weber (SW) formation energy $E_f$ for a defect vacancy line (DVL) (eV per vacant dimer). We report formation energies using either $\mu_{bulk}$, the bulk chemical potential, or $\mu_{dimer}$, the surface chemical potential, as explained in the text.
}
\begin{tabular}{|c|c|c|c|}
\hline
Method & Chemical potential & $E_f$ no strain & $E_f$ 4\% strain \\ \hline
DFT & $\mu_{dimer}$ & 2.28 eV      & -1.09 eV\\
DFT & $\mu_{bulk}$ & -0.11 eV    &   -1.41 eV\\
SW & $\mu_{dimer}$  & 2.85 eV  &  1.56 eV \\
SW & $\mu_{bulk}$  &  0.20  eV  &  -1.01 eV \\
\hline
\end{tabular}

\label{tab:DVL_formation_energy}
\end{table}

\subsection{Displacements relative to the perfect surface}

A recent surface x-ray diffraction (SXRD) experiment \cite{zhou2011atomic} has provided
important data concerning the structure of the Ge-Si(001) DVL. While the experiment cannot distinguish between the effects of the overall compressive strain exerted on the Ge layers and that of the alloying of Ge and Si, our numerical setup allows us to do so: by looking at Si(001) $2\times10$ structures with and without 0.04 biaxial compressive strain, we isolate stress effects. From this, it possible to also identify the separate alloying effects.

\begin{figure*}
	\centering
    \includegraphics[width=18cm]{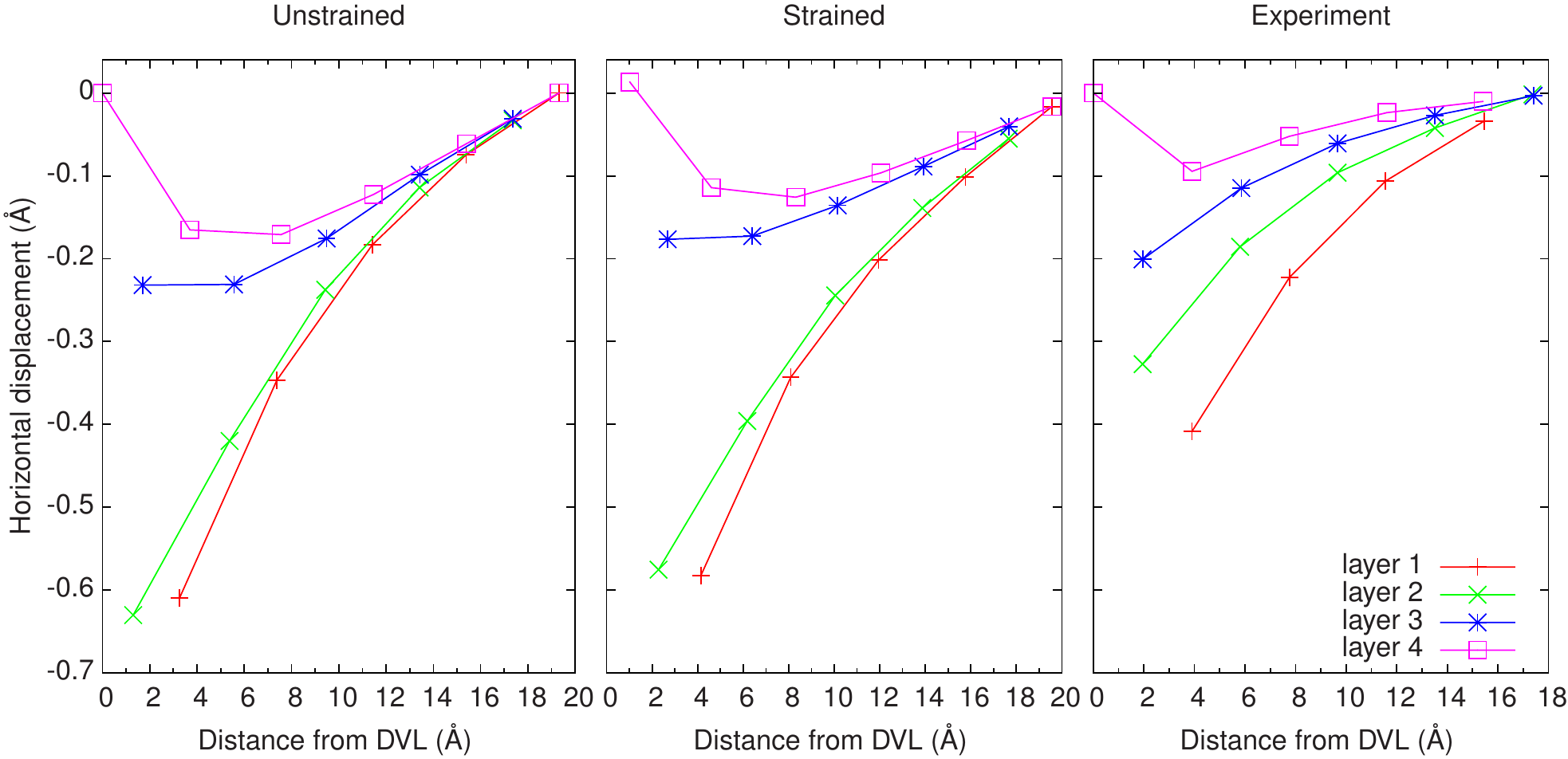}
    \caption{Horizontal displacements measured with respected to the perfect $2\times1$ lattice as a function of distance to the DVL for different depth. Left panel : unstrained sample; middle panel: model with 4\% biaxial strain; right panel: experimental values. Experimental data are taken from Ref. \onlinecite{zhou2011atomic}.} 
\label{fig:x_plot} 
\end{figure*}

\begin{figure*}
	\centering
    \includegraphics[width=18cm]{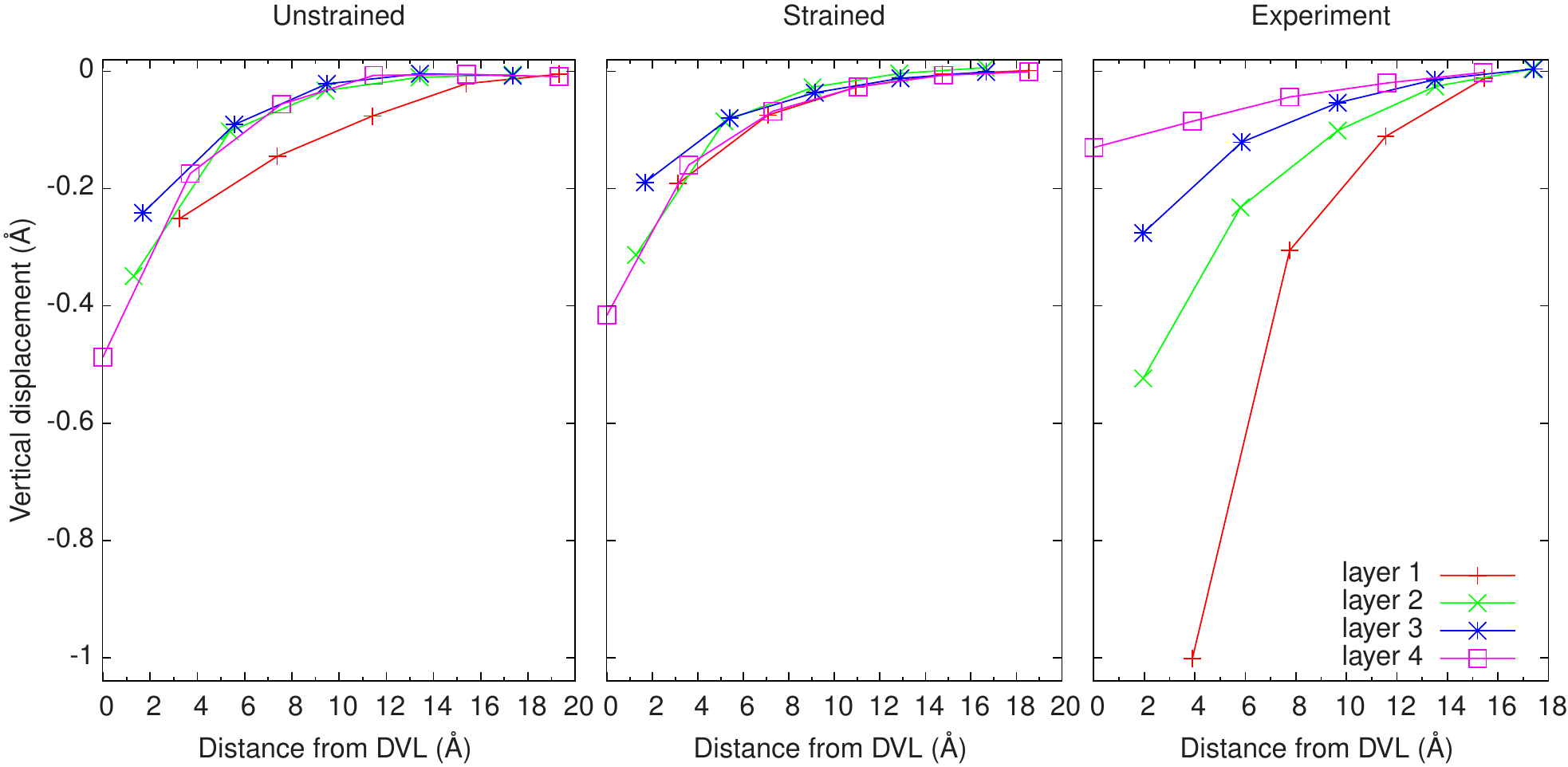}
    \caption{Same as previous figure, but for vertical displacement. Experimental data are taken from  Ref. \onlinecite{zhou2011atomic}.} 
\label{fig:y_plot} 
\end{figure*}

The fully relaxed configuration for a $2\times 10$ DVL, along with atomic vertical and
horizontal displacements with respect to the relaxed perfect $2\times1$ surface, are illustrated
in Figures \ref{fig:map_x} and \ref{fig:map_y} for the surface with no biaxial
strain. Figures~\ref{fig:x_plot}
and \ref{fig:y_plot} present the horizontal and vertical displacement as a function of
the distance to the DVL and the depth below the surface, comparing numerical and
experimental results. While the experiments were performed on a $2\times9$ reconstructed
surface and our calculations on a $2\times10$ reconstructed surface, the weak elastic
deformation at the largest distance from the DVL allows us to do a direct comparison.

Atoms near the DVL are shifted horizontally towards the defect (Fig. \ref{fig:x_plot}). In
the first, second and third layers, these displacements increase monotonously as the
distance to the DVL is reduced and become smaller as a function of depth. In the fourth
layer, we observe a non-monotonous behavior, with no horizontal displacements under the
DVL, a move towards the bulk that peaks about 4~\AA~ away from the DVL and the slow
disappearance of this deformation far from the DVL. The agreement between our
computations and experiments, here, is excellent with a maximum atomic displacement of 0.4~\AA~ in experiment and 0.6~\AA in our models.

Vertically (Fig. \ref{fig:y_plot}), atoms are shifted towards the bulk in the vicinity of
a DVL, a shift that decreases monotonously as the distance for the DVL increases and as a
function of depth. While overall trends are the same for simulations and experiments,
they differ in the details. More precisely, while vertical displacements in our
simulation are of a smaller magnitude in the first layer (up to 0.25 \AA) than in the
experimental case (up to 1~\AA), they are of a comparable magnitude in the second and
third layer and the fourth layer shows larger displacements in our calculation than in
the experiments, close to the DVL (up to 0.5 \AA~ in our calculations and 0.08 \AA~ in
the experiment).

While vertical displacements reported in the experiment dampen to less than 0.05 \AA~ in
the sixth layer, the deformation propagates much deeper in our Si-only sample. A small displacement of 0.05~\AA~ is reached only in the 17th layer. 

To understand these results, we first note that the overall strain has very little impact
on the DVL induced deformation. Horizontal displacements are almost identical for the
strained and unstrained models, and vertical ones are only slightly smaller in the
strained sample (center panel of Fig.~\ref{fig:y_plot}) than in the unstrained sample
(left panel). For instance, the largest displacement in the fourth layer is 0.4 \AA~ for
the strained sample compared to 0.5 \AA~ for the unstrained slab. Differences between
simulations and experiments are therefore mostly due to alloying and size-mismatch disorder between Ge and Si atoms.

Indeed, since the first-neighbor interatomic distance in Ge is about 0.2 \AA~ greater
than in Si, a Ge atom at the surface of a Si wafer requires a smaller horizontal move to
form stable bonds at the DVL. This is what we observe with an experimental displacement
of 0.4~\AA~ versus 0.6~\AA~ for the all-Si simulations. As expected, therefore,
horizontal displacements in the sample with 4 \% biaxial strain (central panel of Figure
\ref{fig:x_plot}) are slightly smaller than those in the unstrained case.

Chemical composition is also responsible for the difference between simulation and
experiment in the vertical displacement. While the strain is uniform as a function of height in our simulation cells, by symmetry, it is depth-dependent in experiments, since the Ge concentration decreases rapidly from 100~\% to 0~\%. This explains the faster dampening as a function of depth observed experimentally, but also the larger reorganization in the top surface near the DVL (Fig.\ref{fig:y_plot}).

\begin{figure}
	\centering
    \includegraphics[width=9cm]{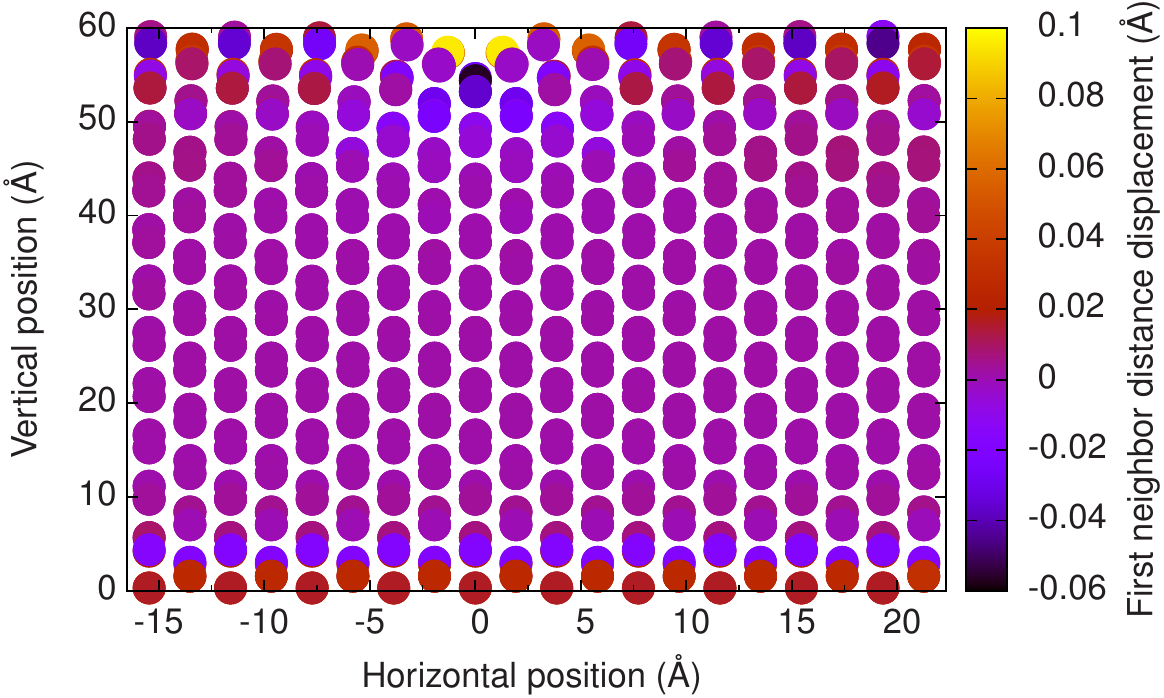}
    \caption{Atomic positions after relaxation in the presence of a DVL. Atoms are color-coded as a function of a change in the first-neighbor distance with respect to that of the non-defective model.} 
\label{fig:map_dist} 
\end{figure}

This point is illustrated in Fig.~ \ref{fig:map_dist}, where the average bond length for each atom is shown (averaged overall all bonds associated with each atom). Near the DVL the atoms in the top two layers have over-extended bonds of up to 0.1~\AA, while the atom
just under the DVL, in the fourth layer, is under considerable compressive strain, with an average bond length of 0.06 \AA~ too short. The presence of Ge in the top layers, coupled with pure Si below the fourth layer, should, in large part, eliminate the compressive strain at the bottom of the DVL.

\section{Energetics of Ge mixing near the DVL}

Both experiments~\cite{zhou2011atomic} and classical Monte Carlo
simulations\cite{nurminen2003reconstruction} show that the Ge concentration profiles
increase with the distance from the DVL, which is interpreted as the consequence of
compressive strain near the DVL. We revisit this issue by studying the energetics of a Ge
atom at various positions in the unstrained $2\times N$ reconstructed Si surface. 

We report the potential energy for the Ge as an adatom, as a an interstitial defect in
the second layer and a substitutional defect in the fourth layer at increasing distances
from the DVL.  For this last system, the energy includes the chemical potential associated with removing a Si-atom from the bulk and placing it as an adatom in a pedestal position. Instead, if one had used the chemical potentials described above, $\mu_{dimer}$ and $\mu_{bulk}$, the potential energy curves for the substitutional defects would have been shifted by 0.29 eV and -0.91 eV, respectively.

\begin{figure}
	\centering
    \includegraphics[width=9cm]{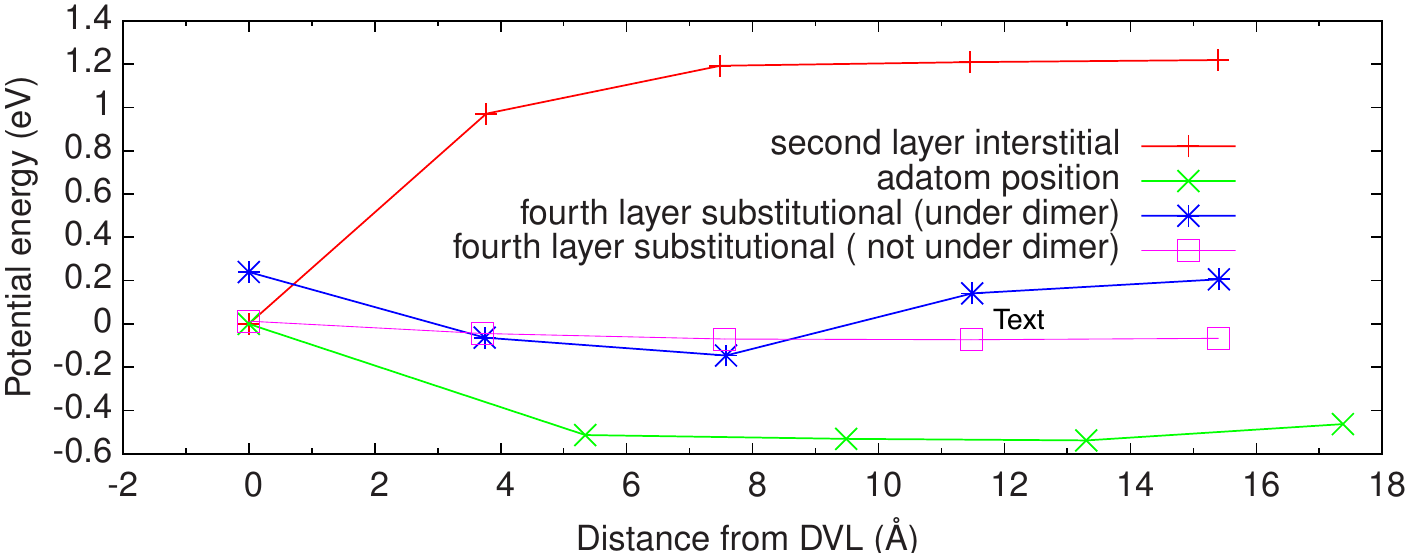}
    \caption{The configurational energy as a function of the position of a Ge defect for a Si slab with a a $2\times N$  reconstructed surface. The Ge is positioned as an adatom at the pedestal position, as an interstitial in the second-layer and as a substitutional atom in the fourth layer (under a dimer row and between dimer rows). In the latter case, the energy plotted includes the chemical potential associated with removing a silicon atom, as discussed in the text.} 
\label{fig:Ge_on_DVL} 
\end{figure}

Above the DVL, the Ge adatom is placed in the middle of the defect; elsewhere, the adatom
is placed in pedestal position. Each configuration is minimized using our QM/MM scheme.
Results are plotted in Fig.~\ref{fig:Ge_on_DVL}. Except over the DVL, where its energy is
0.5~eV larger than in the other positions, the energy of the Ge adatom in pedestal
position is almost independent of its distance to the defect.

The Ge atom in interstitial position is unstable when positioned below the DVL and  spontaneously diffuses above the defect, back in the adatom site (which is why the two points have the same energy in Fig.~\ref{fig:Ge_on_DVL}).  It is also less stable as a second-layer interstitial away from the DVL, with a configurational energy more than 1.7~eV above that of the Ge adatom, except next to the DVL where its energy is 1.5~eV above the Ge adatom. These results, which show that the Ge atoms allow some level of strain relaxation near the DVL, are consistent with the SXRD results discussed in the previous section.

Contrary to adatom and subsurface positions, the energy for a substitutional Ge atom in the fourth layer does not vary monotonously as a function of distance from the DVL. It shows rather a minimum at the third lattice site from the DVL. This effect is strong when the Ge is placed  under the dimer rows, but not in between these rows and it mimics the horizontal displacements in the fourth layer observed  in Fig. ~\ref{fig:x_plot}. A careful inspection of the bond lengths shows that the introduction of this interstitial atom deforms all the bonds in the chain of atoms sitting on the line perpendicular to the DVL, a bit like the compression on a accordion. Bonds near the DVL and the half-distance to its image are stiffer (they move less when introducing a Ge atom), while the bonds in between are softer. The effect is weak in the chain between the dimer rows because the Ge in not as constrained vertically. 

We thus conclude that the presence of a DVL does influence the energetics of Ge intermixing, as predicted and measured by former studies. However, while previous studies stated that Ge concentration decreases monotonously as we approach the DVL, our calculations suggest that concentrations profiles are much more complex, depending on defect type, intermixing depth and the presence (or absence) of a dimer row above Ge sites. While such effects were not reported in Ref. ~\onlinecite{zhou2011atomic}, the very large error bars of the measurements may account for these descrepencies. Ideally, one would want to perform diverse Ge/Si Monte Carlo simulation of intermixing. However, considering the large cost of DFT calculations, it does not seem like a realistic approach at the moment.

\section{Ge/Si intermixing at the surface}

While the previous section identifies some of the most stable sites for a Ge atom in the
top layers of a Si slab with a DVL, the sampling of these positions is constrained by the
kinetics of Ge mixing. Uberuaga \emph{et al.}'s seminal work \cite{uberuaga2000diffusion}
suggests a path from the surface to the third layer with a maximum barrier of 1.3 eV, an
energy that would allow mixing in the top three layers at temperatures well below 600 K.

However, the identified path from the third to the fourth atomic layer shows a 2.3 eV
barrier relative to the global minimum energy state, leading to a metastable state 1.75
eV above the global minimum. On experimental time scale, this particular path limits the
Ge mixing to temperatures above 773 K, suggesting a kinetic barrier between the third and
fourth layers. Yet, using the results presented in the previous section and
thermodynamical computations \cite{uberuaga2000diffusion}, recalculated with T=600 K, we
predict that the population of Ge in the fourth layer, at thermodynamical equilibrium
should be close to two percent. 

Previous experiments have characterized the top two layers of Ge/Si, but no definitive
data is available, to our knowledge, concerning deeper layers: results between 573~K and
773~K are ambiguous concerning this issue \cite{nakajima1999direct}. Since thermodynamics
suggests that Ge should be present in the fourth layer, we select to extend the search
for kinetic pathways initially performed by Uberuaga \emph{et al}, using using an
open-ended technique, ARTn, in order to search for kinetically accessible pathways to these deep layers
\cite{PhysRevLett.77.4358,malek2000dynamics,machado2011optimized,mousseau2012activation}.

We first sample the energy landscape of the Ge adatom using the QM/MM slab with 24 atoms per layer and one Ge adatom, generating more than 100 metastable states and associated transition states. We confirm that the pedestal position is the most stable configuration. The energy landscape is very rugged, however, with many local minima where the adatom sits above the Si dimers linked by saddle points with an energy barrier of 0.05-0.3 eV. Yet, the computed diffusion barrier along the dimer row is 0.60 eV, in good agreement with previous studies.
 
\begin{table}
\centering
\caption{Left column: the potential energy of metastable (bold font) and transition states (normal font) that permit the diffusion of a Ge atom from the third to the fourth layer of a Si surface. Middle column: atomistic configuration of each state as seen from the X direction; right column: atomistic configuration as seen from the Z direction. Si are represented by small green spheres and Ge by large blue spheres.}
\begin{tabular}{ccc}
   Energy (eV) & X direction & Z direction \\ \hline
  & & \\
	  \begin{tabular}{c} \textbf{M1} \\ \textbf{1.03} \end{tabular}   &   
      \includegraphics[width=2.7cm]{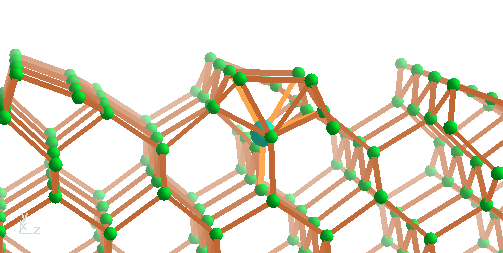}         & 
		  \includegraphics[width=2.7cm]{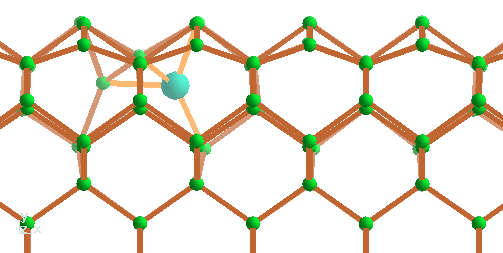}                 \\ 
		  
		\begin{tabular}{c} S1 \\ 1.55  \end{tabular}         &    
        \includegraphics[width=2.7cm]{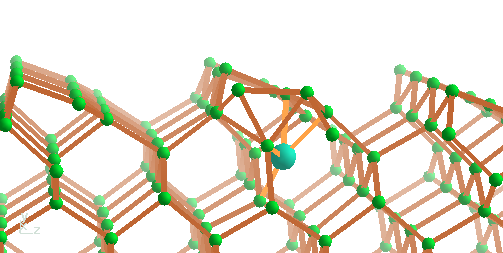}     & 
			\includegraphics[width=2.7cm]{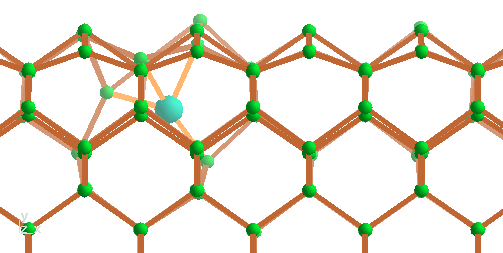}                  \\
			
		\begin{tabular}{c} \textbf{M2} \\ \textbf{1.14} \end{tabular}   &     
        \includegraphics[width=2.7cm]{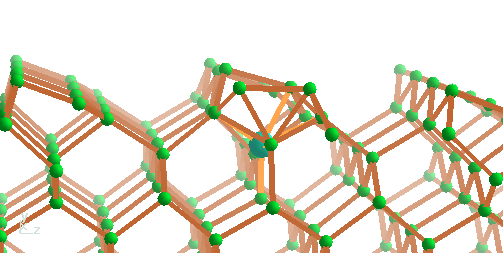}             &
			\includegraphics[width=2.7cm]{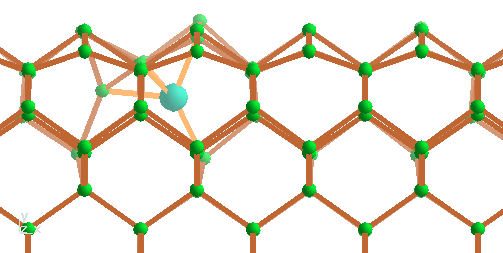}                  \\
			
		 \begin{tabular}{c} S2 \\ 1.43 \end{tabular}   &  
         \includegraphics[width=2.7cm]{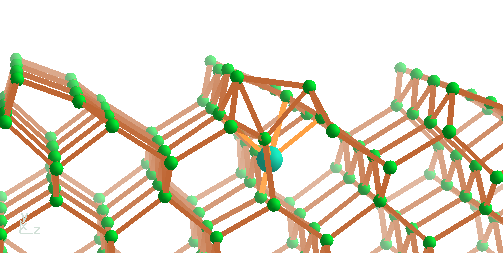}             &  
			 \includegraphics[width=2.7cm]{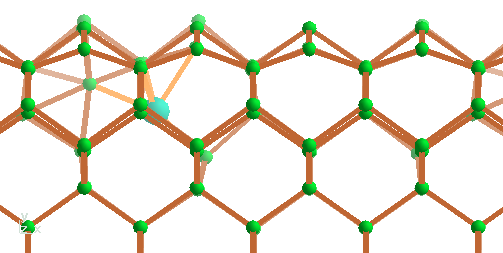}                  \\
			 
		 \begin{tabular}{c} \textbf{M3} \\ \textbf{1.33} \end{tabular}  &   
         \includegraphics[width=2.7cm]{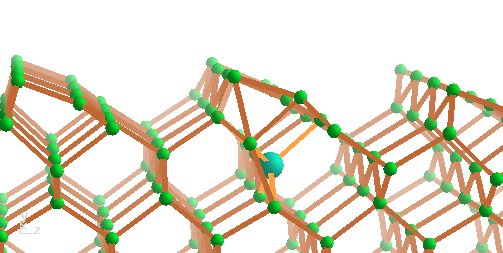}             & 
			 \includegraphics[width=2.7cm]{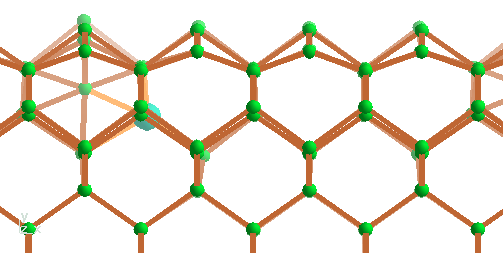}                  \\
			 
	 \begin{tabular}{c} S3 \\1.82 \end{tabular}  &       
     \includegraphics[width=2.7cm]{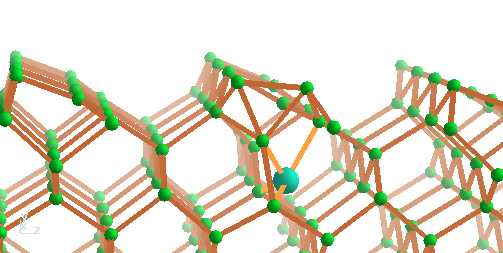}             & 
		 \includegraphics[width=2.7cm]{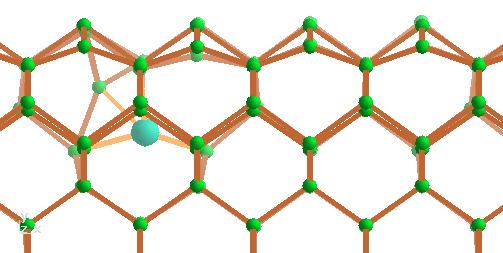}                  \\
		 
	 \begin{tabular}{c} \textbf{M4} \\ \textbf{1.63} \end{tabular}    &      
     \includegraphics[width=2.7cm]{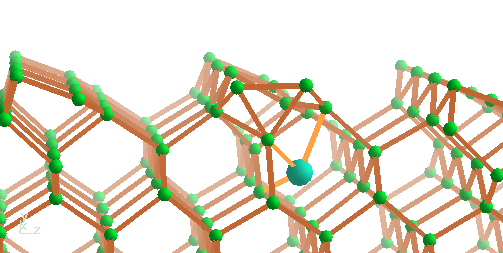}             &
		 \includegraphics[width=2.7cm]{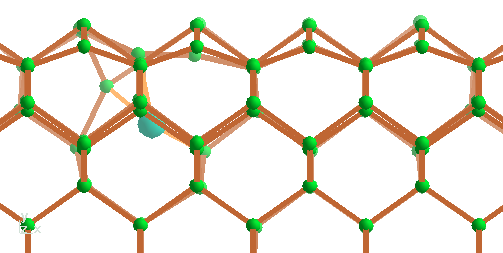}                  \\
		 
	  \begin{tabular}{c} S4 \\ 1.68 \end{tabular}  &       
      \includegraphics[width=2.7cm]{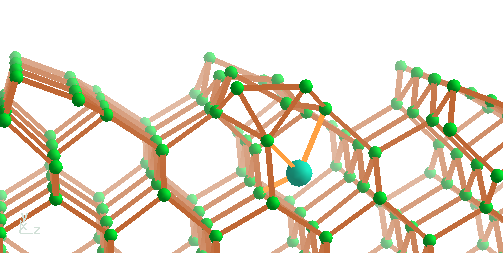}             & 
		  \includegraphics[width=2.7cm]{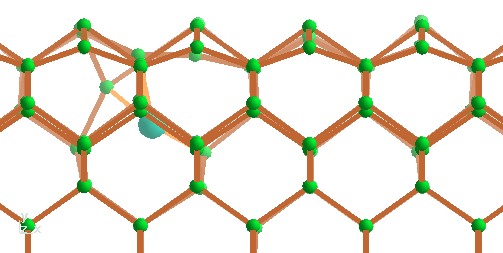}                 \\
		  
	\begin{tabular}{c} \textbf{M5} \\ \textbf{1.38} \end{tabular}  &        
    \includegraphics[width=2.7cm]{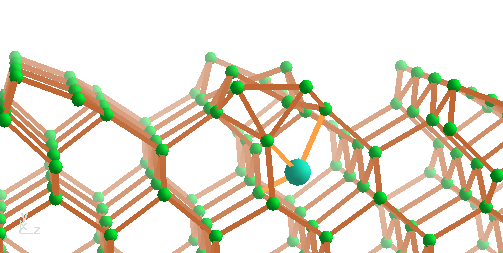}             & 
		\includegraphics[width=2.7cm]{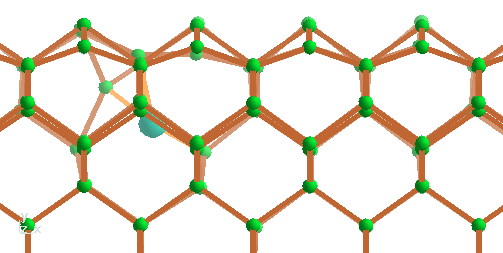}                 \\
		
	 \begin{tabular}{c} S5 \\ 1.82 \end{tabular}  &      
     \includegraphics[width=2.7cm]{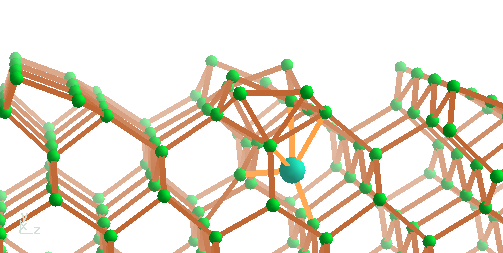}             & 
		 \includegraphics[width=2.7cm]{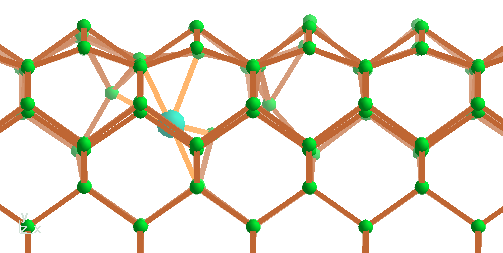}                   \\
		 
	  \begin{tabular}{c} \textbf{M6} \\ \textbf{1.63} \end{tabular}   &     
      \includegraphics[width=2.7cm]{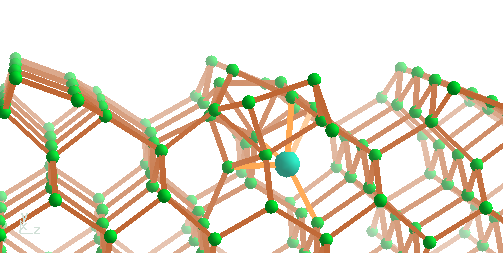}             & 
		  \includegraphics[width=2.7cm]{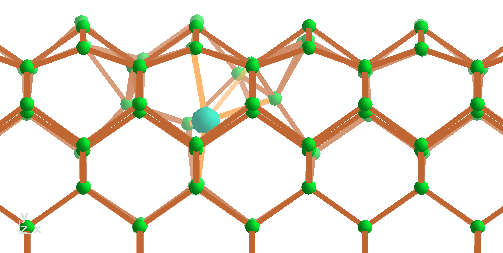}                   \\
		  
	 \begin{tabular}{c} S6 \\ 1.82 \end{tabular}  &     
     \includegraphics[width=2.7cm]{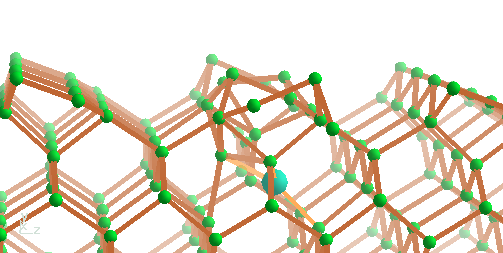}             &  
		 \includegraphics[width=2.7cm]{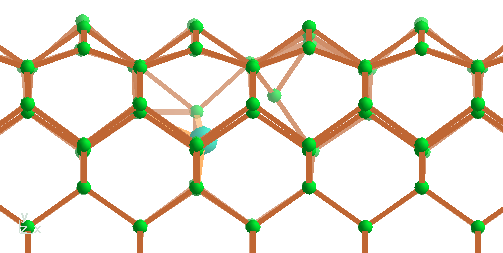}                   \\
		 
	 \begin{tabular}{c} \textbf{M7} \\ \textbf{0.70} \end{tabular}   &     
     \includegraphics[width=2.7cm]{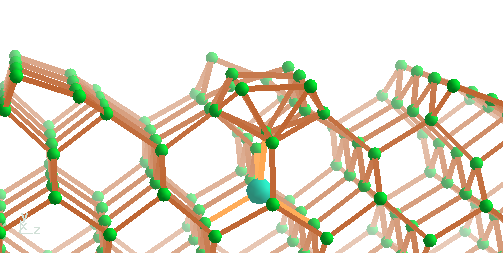}             &  
		 \includegraphics[width=2.7cm]{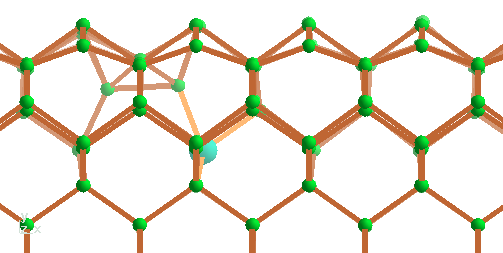}           \\ \hline  
\end{tabular}
\label{tab:Ge_interdiff}
\end{table}

We then place a Ge atom in a Si-Ge dumbbell position in the third layer (firt row in table \ref{tab:Ge_interdiff}), in a
configuration corresponding to the last accessible state at less than 600K found in
Uberuaga \emph{et al.}'s study \cite{uberuaga2000diffusion}. According to this group,
this state is 0.3 eV above that of the Ge in a pedestal position.

Relaxing our configuration with QM/MM, we find a significantly higher energy: 1.03 eV
above that of the Ge in a pedestal position. This difference in energy can be explained
in part by the surface reconstruction. Here, the surface dimers are in a 2$\times$1 while
there are in a 2$\times$2 reconstruction in Ref. ~\onlinecite{uberuaga2000diffusion},
justifying a 0.3~eV spread. The rest must come from the choice in the DFT calculation and
the box set-up.

Running an ART nouveau search from this state, and focusing on migration pathways leading
away from the surface, we find a multistep diffusion path from the third to the fourth
layer with a 1.82 eV activation barrier as computed from the pedestal position, lower
than the 2.3 eV previously found. The final state is also lower in energy, at 0.67~eV,
compared with 2~eV found in Ref.~\onlinecite{uberuaga2000diffusion}.

The details of the path from the third to the fourth layer are given in Table
\ref{tab:Ge_interdiff} and are similar to the diffusion mechanism described in Ref.
~\onlinecite{caliste2007germanium}. Minima 1 through 5 are associated with the migration from a Ge-Si dumbbell configuration in the third layer to an Ge-interstitial hexagonal configuration between the third and fourth layer. Saddle 5 corresponds to the diffusion of Ge from this hexagonal site to a distorted Si-Ge dumbbell in the fourth layer, (minimum 6). Saddle 6 leads to a substitutional Ge atom in layer 4 and an Si-Si dumbbell in layer 3 (minimum 7). 

Using harmonic transition state theory with a standard pre-exponential for surface diffusion of 100 THz and a barrier of 1.8 eV, diffusion from the third to the fourth layer would be limited to once every 13 seconds at 600 K, which is coherent with deposition speeds of the order of ML/min. In principle, this means intermixing should be kinetically feasible at 600K. Overall, these theoretical findings warrant further experimental investigations, since ambiguity remains concerning inter-diffusion at that temperature.

It is interesting to note that elastic effects are not uniform along the diffusion pathway but are particularly important at the transition states below the surface. Pure QM calculations reproduce closely, within 0.05 eV, the energy minima, when compared with QM/MM, but overestimate by as much as 0.5 eV the transition states. This apparent asymmetry between minimum and activated states can be explained by the larger lattice deformation observed at the transition that increases the elastic impact on the total energy. The importance of taking into account long-range elastic deformation through QM/MM is also increased as the Ge diffuses far from the surface and closer to the bottom QM layer. It is therefore crucial, for the right kinetics, to take elastic effects correctly into account.


\section{Conclusion}

This work is concerned with understanding the onset of Ge mixing in Si(100) using a quantum mechanical/molecular mechanics approach based on the BigDFT package. 

Focusing on the role of strain, we characterize the structural and thermodynamical impacts of creating a dimer vacancy line (DVL) at the surface of a pure Si(100) slab at zero and 4 \% compressive strain. This allows us to show the importance of taking into accounts deep elastic effects to evaluate correctly the elastic energy associated with the surface deformation. Computing the formation energy, we also show that previous calculations using the bulk chemical potential as reference lead to predicting the stability of the DVL in the pure unstrained Si(100), a defect that is not observed experimentally. It is necessary to use a surface chemical potential to recover the proper thermodynamics.

Comparing the lattice deformation with recent experimental data, we also identify the role of the strain vs. chemical disorder, particularly near the DVL, as a new pathway  leading from the surface to the fourth layer with a lower activation barrier (1.82 eV) than that found in previous studies (2.3 eV). This is coherent with experiments that suggest, although ambiguously, that inter-diffusion can occur at temperatures lower than 773K. 

These findings should raise new experimental investigations. Besides, the results above demonstrate the importance of taking into account elastic effects when computing the structural properties of semiconductors surface. Elastic deformation play an important role in this structure and should be taken into account to properly describe surface structures. Furthermore, as already shown for binary systems \cite{machado2012tunable}, our calculations stress the critical importance of the choice of chemical potential when computing formation energies of surface structures. While a convenient choice, the use of the bulk chemical potential, results in predicting that DVLs spontaneously appear on unstrained Si(001), choosing the surface dimers binding energy as a chemical potential results in predicting that compressive strain is necessary for the DVL to appear. These results remind us of the ambiguities involved in computing formation free energies in a grand canonical ensemble. If one wants to determine with precision the adequate chemical potential, large simulations in the canonical ensemble would be required.

\begin{acknowledgments}
We are grateful to Dr. G. Renaud and T. Zhou for insightful discussions. 
LKB acknowledges Dr. T. Albaret, Dr. L. Genovese and Dr. D. Caliste for key discussions during his three month stay in Grenoble. 
We thank Calcul Qu\'{e}bec and GENCI (gen6107) for generous allocation of computer resources. 

This work benefited from the financial
support of NanoQu\'{e}bec, the Fonds qu\'{e}b\'{e}cois de recherche sur la
nature et les technologies, the Natural Science and Engineering Research
council of Canada, the Canada Research Chair Foundation, and the fondation Nanosciences.
Some of LKB's work was supported by the Center for Defect Physics,
an Energy Frontier Research Center funded by the U.S. Department of Energy,
Office of Science, Office of Basic Energy Sciences.
\end{acknowledgments}

\bibliography{Biblio_Si_Ge}

\end{document}